\documentstyle[12pt]{article}

\newcommand{\be}{\begin{equation}}
\newcommand{\ee}{\end{equation}}

\begin{document}

\begin{titlepage}
\centerline{\large\bf INSTITUTE OF THEORETICAL AND EXPERIMENTAL PHYSICS}
\vspace{7cm}
\centerline{\large\bf B.O.Kerbikov and Yu.A.Simonov}
\vspace{2cm}
 \centerline{\large\bf
PATH INTEGRAL BOSONIZATION OF THREE FLAVOUR
}
\vspace{5mm} \centerline{\large\bf
 QUARK
DYNAMICS}
\vspace{5cm}

\centerline{\large\bf Moscow - 1995}
\vspace{1cm}
$~~$

\end{titlepage}

\newpage
$~~$
\vspace{60mm}

\centerline{\large\bf A b s t r a c t}
\vspace{5mm}
\large

Chiral symmetry breaking due to instanton-produced fermion zero modes
in the confining vacuum is considered. Zero modes provide 'tHooft-type
determinant quark interaction, which is bosonized by introduction of
auxiliary fields. For three flavours this procedure becomes nontrivial
and a method is suggested which allows to derive effective
Lagrangian for massless Nambu--Goldstone modes.
%\end{abstract}

%\newpage
 \setcounter{equation}{0}
\renewcommand{\theequation}{1.\arabic{equation}}

\newpage

\section{Introduction}

Recently effective  action has been obtained for a theory with
confining background superimposed on instantons [1]. This is an
extension of earlier investigations of the  pure instanton vacuum
[2-4]. The aim of [1]  and of the present letter is to formulate a
model of the QCD vacuum with properties of confinement and chiral
symmetry breaking (CSB) at the same time. However in [1] only one--
and two-- flavour $(N_f=1,2)$ cases have been considered. It is known
that bosonization becomes really nontrivial for $N_f>2$ [5].
In the present paper we perform explicit path integral bosonization
of  quark dynamics outlined in [1] for $N_f=3$ case.

First we briefly remind the basic features of the QCD model
developed in [1]. The driving idea is to consider chiral and
confinement effects as interconnected. To this end the QCD vacuum is
assumed to consist of instantons plus confining configurations. Thus
the starting point is the ansatz for the vacuum
\be
A_{\mu} = \sum^N_{i=1} A_{\mu}^{i}+B_{\mu}~,~~~N=N_++N_-~,
\ee
 where $A^i_{\mu}$ is the field of $i$-th (anti)instanton ($N$ is the
 total number of instantons and antiinstantons in the 4--volume $V$),
 $B$ is the background field which ensures confinement
 (i.e.correlators $F_{\mu\nu}(B)$ yield nonzero srtring tension).
 Instantons through fermion zero modes give rise to CSB.

 Confining background on one hand modifies the instanton density [6],
 and on the other it interplays with instantons and hence modifies
 chiral effects [1].

 As shown in [1] the gauge invariant partition function of quarks in
 the field (1.1) in the limit of zero quark masses can be written in
 a form similar to that of the pure instanton gas [4]:
 \be
 Z=\int D\psi D\psi^+D{\mu}(B)\int dg_+dg_- exp W~,
 \ee
 \be
 W=-N_+lng_+-N_-lng_-+\int d^4u\{\psi^+_f(i\hat{D})\psi_f+g_+
 detJ_++g_- det J_-\}~,
 \ee
 where the action of the gluonic field is included in $D{\mu}(B)$,
 the integration over $g_+$ and $g_-$ is  a remnant of the inverse
 Laplace transform introduced in order to rewrite the partition
 function in the exponential form [4],
 $f=1,2,...,N_f$ is the flavour index, $N_+$ and $N_-$ were defined
 in Eq.(1.1), $\hat{D}\equiv \hat{D}(B_{\mu})$ is the covariant
 derivative  corresponding to the field $B_{\mu}$, the $N_f\times
 N_f$ matrices $J_{\pm}$ are related to $2N_f$ fermion  vertices
 $Y_{\pm}$ via the relations
 \be
 Y_{\pm}=\int d^4u det J_{\pm}(u)~,
 \ee
 \be
 (J_{\pm}(u))_{fg}=\int dx dy \psi^+_f(x)\frac{1}{2}
 (1\mp\gamma_5)K(x,y,u)\psi_g(y)~,
 \ee
 \be
 K(x,y,u) = i\hat{D}\bar{\varphi}(x-u)\Phi(x,u,y)
 \bar{\varphi}^+(y-u)i\hat{D}~,
 \ee
 where $\bar{\varphi}(x)$ is the instanton zero--mode solution:
  \be
  \bar{\varphi}(x) =\frac{1}{\pi} \frac{\rho}{(x^2+\rho^2)^{3/2} }
  ~~~\frac{x_{\mu}\gamma_{\mu}}{\sqrt{x^2}}
  \ee
 and $\Phi(x,u,y)\equiv \Phi(x,u)\Phi(u,y)$ is a product of parallel
 transporters
  \be
  \Phi(x,u)= P exp ~(ig \int^x_u B_{\mu} dz_{\mu}).
  \ee
 Note that $J_{\pm}$ and $W$ are gauge invariant due to the presence
 of long derivatives $\hat{D}$  and parallel transporters in (1.6).

 Expressions (1.2-1.3) for the partition function are the starting
 point for the rest of the paper. Our goal is to derive an effective
 action by integrating over the fermionic degrees of freedom. The
 most efficient way to do this is to introduce auxiliary fields
 (bosonization), then perform the integration over fermions exactly,
 and then apply the stationary--phase approximation to the integral
 over the auxiliary field. As a result one obtains the effective
 action for Nambu--Goldstone modes.

 \setcounter{equation}{0}
\renewcommand{\theequation}{2.\arabic{equation}}

 \section{The case of two flavours}

The usual path integral bosonization is based on the
Hubbard--Stratonovich transformation [7].
For the quark dynamics under consideration this procedure has been
performed in [4] and [1]. However it faces problems for $N_f>2$, and
more  sophisticated manipulations are needed. In this section we
shall reconsider the $N_f=2$ case (see Ref.[1]) within the framework
of the approach applicable for arbitrary $N_f$. The basics of the
bosonization techniques which will be used here have been developed
in [8,9].

Consider the partition function (1.2-1.3) for $N_f=2$. The general
relation, which holds for any $2\times 2$ matrix, allows  to write
\be
\det J_{\pm}=\frac{1}{2}[(tr J_{\pm})^2-tr J^2_{\pm}]~.
\ee
Expanding $J_{\pm}$ over a complete set of three Pauli matrices
$\tau^a(a=1,2,3)$ plus a matrix $\tau^0=-i\hat{1}$ yields
\be
J_{\pm}=\sum^3_{a=0}\tau^a c^{\pm}_a~,~~c^{\pm}_{a=1,2,3}=\frac{1}{2}
tr (\tau^aJ_{\pm})~, ~~ c_0^{\pm}=-\frac{1}{2} tr(\tau^0 J_{\pm}).
\ee
Then by virtue of (2.1) we get
\be
det J_{\pm} = -\sum^3_{a=0}(c^{\pm}_a)^2~.
\ee

Next we split out from the partition function $Z$ given by Eqs.
(1.2-1.3) the  following  part:
\be
\tilde{Z}=\int D\psi D\psi^+ exp \{ \int d^4 u
[\psi^+_f(i\hat{D})\psi_f+g_+ det J_+(u) +g_- detJ_- (u)]\}
\ee
$$\equiv \int
D\psi D\psi^+ exp \tilde{W}.
$$
This expression can be written in the following equivalent form
(index $a$ runs from 0 to 3):
$$
\tilde{Z}=\int D\psi D\psi^+\prod_a D\sigma^+_a\prod_a D\sigma^-_a
\delta(\sigma^+_a-g_+c^+_a) \delta(\sigma^-_a-g_-c^-_a) exp
(\tilde{W})=
$$
$$
=\int D\psi D\psi^+\prod_a D\sigma^+_a\prod_a D\sigma^-_a
\prod_aDM^+_a\prod_a DM^-_a exp (\tilde{W})\times
$$
\be
\times exp\{ i\int d^4 u
[M^+_a(\sigma^+_a-g_+c^+_a)+M^-_a(\sigma^-_a-g_-c^-_a)]\}~.
\ee

At the next step we integrate out in (2.5) the  quark fields $\psi$
and $\psi^+$, i.e. perform the integration
\be
\tilde{Z}_{\psi}=\int D\psi D\psi^+
exp
\{ i\int d^4 u
[\psi^+_f\hat{D}\psi_f-g_+M^+_ac^+_a-g_-M^-_ac^-_a)]\}~,
\ee
where according to (2.2) and (1.5) we have
$$
M_a^{\pm}c_a^{\pm}=\frac{1}{2}\sum_a M_a^{\pm}\sum_{fg}(\tau^a)_{fg}
J^{\pm}_{gf}=
$$
\be
=\int d^4x d^4y\psi^+_g(x)[\frac{1}{2}
M^{\pm}(u)(\frac{1\mp\gamma_5}{2})K(x,y,u)]\psi_f(y)~,
\ee
with $M^{\pm}(u)\equiv \{
 M^{\pm}_{a>0}(u)\tau^{a>0},~-M^{\pm}_0\tau^0\}$.  The integration yields:
\be
\tilde{Z}_{\psi}=-exp (\ln~det X)
\ee
\be
 X=
 \left( \begin{array}{ll} -i({D}_4-i
\vec{\sigma}\vec{D})&\frac{i}{2}g_+M^+K\\ &\\ \frac{i}{2}g_-M^-K&
-i(D_4+i\vec{\sigma}\vec{D})
\end{array}
\right)\\ .
\ee
Next one performs integration over the fields $\sigma^{\pm}_a$
\be
\tilde{Z}_{\sigma}=\int\prod_a D\sigma^+_a\prod_a D\sigma^-_a  exp
\{ \int d^4 u
[g_+ det J_++g_-
det
J_-+iM^+_a\sigma^+_a+iM_a^-\sigma_a^-]\}~.
\ee

Making use of (2.3) and (2.5) one has
\be
g_{\pm} \det J_{\pm}=-\frac{(\sigma_a^{\pm})^2}{g_{\pm}}~,
\ee
where summation over $a$ is implied. Therefore integral (2.10) is
Gaussian with the result
\be
\tilde{Z}_{\sigma}= exp \{ -\frac{1}{4}\int d^4 u
[g_+(M^+_a)^2+g_-(M^-_a)^2]\}~.
\ee

With the identification of the fields $M^{\pm}_a$ with $L_a$ and
$R_a$ of Refs. [1,4]
\be
L_a=\frac{i}{2}M^+_a~,~~R_a=\frac{i}{2}M^-_a~.
\ee
the final results for fully bosonic effective action reads
$$
exp\{-S(L,R)\}=\int dg_+ dg_- exp\{-N_+ ln g_+-N_-lng_-+
$$
\be
 +\int d^4u[g_+L_a^2(u)+g_-R^2_a(u)]+ln ~det~ X~,
 \ee
\be
X=\left( \begin{array}{ll}
-i({D}_4-i\vec{\sigma}\vec{D})&g_+{L}K\\
 &\\
 g_-{R}K&
-i({D}_4+i\vec{\sigma}\vec{D})
\end{array}
\right)\\ .
\ee

 For further purposes it is instructive to rewrite the term
 $ln~det~X$ in (2.14-2.15) in a slightly different form. In
 (2.14-2.15) the determinant in Dirac space is written down
 explicitly. Instead let this step be implied in $det~X$ notation.
 Then making use of the identity $\ln\det~X=tr\ln X$ one can recast
 $\ln\det X$ in (2.14) into the form
 \be
 \ln\det X=
 tr\ln\{i\hat{D}+g_+(\frac{1-\gamma_5}{2})LK+g_-(\frac{1+\gamma_5}{2}) RK\}~,
 \ee

 This form will be used in Section 4 in order to define the chiral
 effective Lagrangian.

 We have retrieved the results obtained in [1]. The reader is
 referred to [1] for a discussion of the physics behind Eqs.
 (2.14-2.16). For $N_f=2$ our present derivation is more cumbersome
 than the Hubbard--Stratonovich transformation used in [1]. However
 our goal was   to display the algorithm applicable for $N_f>2$ case.

 \setcounter{equation}{0}
\renewcommand{\theequation}{3.\arabic{equation}}

 \section{Bosonization of Three Flavours Dynamics}

 Again our starting point is the  partition function given by
 Eqs.(1.2-1.3), and our task is the integration over fermionic
 degrees of freedom.

 For the case $N_f=3$ relations (2.1-2.3) are replaced by
 \be
 \det J_{\pm}=\frac{1}{6}(tr J_{\pm})^3-\frac{1}{2}(tr J_{\pm})(tr
 J^2_{\pm})+\frac{1}{3} tr J^3_{\pm}~,
 \ee
 \be
 J_{\pm}=\sum^8_{a=0}\lambda^a c^{\pm}_a~,~~c_{a>0}=\frac{1}{2} tr
 \lambda^a J_{\pm}~,~~c_0=\frac{1}{3} tr (\lambda^0J_{\pm})~,
 \ee
 \be
 \det
 J_{\pm}=(c^{\pm}_0)^3-\frac{5}{2}c^{\pm}_0\sum^8_{a=1}(c_a^{\pm})^2+
 \frac{1}{3}
 \sum^8_{a,b,d=1}
 tr(\lambda^a\lambda^b\lambda^d)c_a^{\pm}c_b^{\pm}c^{\pm}_d~.
 \ee
 Here $\lambda^a,~~a=1,2,..,8$ are the SU(3) Gell-Mann matrices,
 $\lambda^{a=0}=\hat{1}$.

 Next one again proceeds as in equations (2.4-2.6) with the only
 difference that the index $a$ now runs from 0 to 8. In (2.7) the
 $\tau_a$ matrices are replaced by $\lambda^a$, so that now
 $M^{\pm}(u)\equiv M^{\pm}_a(u)\lambda^a$.
In line with (2.8-2.9) and (2.16) integration over $D\psi D\psi^+$
yields
 \be
  \tilde{Z}_{\psi} = exp \{ tr \ln [ i
\hat{D}+g_+(\frac{1-\gamma_5}{2}) LK +g_-
(\frac{1+\gamma_5}{2})RK]\}~.
\ee
Here $L=\frac{i}{2}M^+=L_a\lambda^a~, ~~R=\frac{i}{2}M^-=R_a\lambda^a$,
so that (3.4)
contains 18 bosonic fields $L_a,R_a,a=0,1,...,8$.

The next step is the integration over $\sigma^{\pm}_a$ and that
is where the real
 difference from the $N_f=2$ case starts. The integral analogous to
 (2.10) reads
\begin{eqnarray}
 \tilde{Z}_{\sigma} = \int\prod_a D\sigma_a \exp
  \{ - \int d^4 u [-\frac{1}{g^2}
 (\sigma_0^3-\frac{5}{2}\sigma_0\sum^8_{a=1}
 \sigma_a^2+\frac{1}{3}\sum^8_{a,b,d=1}
 tr(\lambda^a\lambda^b\lambda^d)\sigma_a\sigma_b\sigma_d)-\\
 \nonumber
 -i\sum^8_{a=0}M_a\sigma_a]\}\equiv \int \prod_a D\sigma_a \exp
\{ S(\sigma_a)\}.
 \end{eqnarray}
  In (3.5) we have suppressed the $(\pm)$
indices, so that additional summation over $(\pm)$ is implied. The
$N_f=2$ integral (2.10) was Gaussian and therefore it could be
calculated both exactly and by stationary phase method yielding the
same results.Exact analytical calculation of (3.5) is not possible.
 Therefore one has to resort to the stationary phase, or the steepest
 descent methods.  The corresponding conditions which define the
 $\sigma_a=f(\{M_a\})$ solutions have the form
 $$ \frac{\delta
 S}{\delta
 \sigma_0}=3\sigma^2_0-\frac{5}{2}\sum^8_{a=1}\sigma^2_a+ig^2 M_0=0
 $$
 \be
 \frac{\delta S}{\delta
 \sigma_a}=-5\sigma_0\sigma_a+\frac{1}{3}\sum^8_{b,d=1}
 tr (\lambda^a\lambda^b\lambda^d)\sigma_b\sigma_d+ig^2 M_a=0~,~~
     a=1,2,...,8~.
 \ee

The principal question is the existence of a contour along which the
integral (3.5) converges,
     so that the steepest descent method yields reliable results. As
     a toy model consider the case $\sigma_0\neq 0$, $\sigma_a=0$,
     $a=1,2,...,8$ (for   $N_f=2$ similar situation was investigated
     in [1,4]). Then (3.5) reduces to
     \be
     \tilde{\tilde{Z}}_{\sigma}=\int D\sigma exp \{ -\int d^4 u
     (-\frac{\sigma^3}{g^2}- i M\sigma)\}
     \ee
     with $\sigma\equiv \sigma_0$. Replacing $M=-2iL,~g=i\kappa$, we
     get
     \be
     \tilde{\tilde{Z}}_{\sigma}=\int D\sigma exp \{ \int d^4 u
     (2L\sigma -\frac{\sigma^3}{\kappa^2})\}~.
     \ee
     This is an Airy type integral [10]. In the Appendix we show how
     to choose the correct contour and how to calculate this integral
     by the steepest descent method. $Up$ to unimportant overall
     factor the result is
     \be
     \tilde{\tilde{Z}}_{\sigma}= exp \{-\frac{4}{3} \sqrt{\frac{2}{3}}
     \kappa \int d^4 u L^{3/2} (u)\}~.
     \ee
     This result is reminisced of the stationary phase
     condition for Landau magnetisation function of the
     infinite range Ising model [7].

 \setcounter{equation}{0}
\renewcommand{\theequation}{4.\arabic{equation}}

 \section{The gap equation}

 For the toy model under consideration the partition function
     (1.2-1.3) reads
     \be
     Z\propto \int DL\int d\kappa exp \{ -N\ln
     \kappa-\frac{4}{3}\sqrt{\frac{2}{3}}\kappa \int d^4 u L^{3/2} +
     tr \ln (-D^2+\kappa^2L^2K^2)\}~.
     \ee
     The steepest descent integration yields
     $$
     -2\sqrt{\frac{2}{3}} \kappa VL^{1/2}+2\kappa^2L~ tr
     (\frac{K^2}{-D^2+\kappa^2L^2K^2})=0~,
     $$
     \be
     -\frac{N}{\kappa}-\frac{4}{3}\sqrt{\frac{2}{3}}VL^{3/2}+2\kappa
     L^2 tr (\frac{K^2}{-D^2+\kappa^2L^2K^2})=0~.
     \ee
     From (4.2) we find
     \be
     \kappa^2L^2 tr (\frac{K^2}{-D^2+\kappa^2L^2K^2})=\frac{3}{2}N~.
     \ee
      With the identification for the chiral mass operator
      $m\equiv \kappa LK$, and performing on the l.h.s. of
     (4.3) trivial summation over flavour indices one gets
     the same form of the gap equation as for $N_f=1,2$ in
     [1]
     \be
     tr(\frac{m^2}{-D^2+m^2})=\frac{N}{2}~.
     \ee

 \setcounter{equation}{0}
\renewcommand{\theequation}{5.\arabic{equation}}

 \section{The effective chiral lagrangian}

     We turn now to the final topic, namely to the constuction of the
     effective chiral Lagrangian for $N_f=3$. Such a Lagrangian
     corresponds to an octet of Goldstone bosons
     $\pi^0,\pi^{\pm},K^{\pm},K^0,\bar{K}^0,\eta$ which are coupled
     by the fermionic loop. The starting point is Eq.(3.4) in which
     we parametrize bosonic fields $L$ and $R$ in the following form
     [4]
     \be
     L=L_a\lambda^a= i\rho(\frac{1+\xi+\mu}{2})UV,~
     R=R_a\lambda^a= i\rho(\frac{1+\xi+\mu}{2})VU^+,~
     a=0,1,...,8,
     \ee
     \be
     U=\exp\{\frac{1}{2}
     i\pi_b\lambda^b\}~,V=exp\{\frac{1}{2}i\varepsilon_b
     \lambda^b\}~,~~ b=1,2,...,8.
     \ee
     This is a general parametrization which expresses 18 bosonic
     fields $L_a, R_a$ in terms of 18 fields $\xi, \mu, \pi_b,
     \varepsilon_b$(note that in (5.1) summation is from 0 to 8,
     while in (5.2) from 1 to 8). Insertion of (5.1)-(5.2) into (3.4)
     yields
     \be
     \tilde{Z}_{\psi}=\exp\{ tr\ln
     [i\hat{D}+i(1+\xi+\mu)UVE^++i(1+\xi-\mu)VU^+E^-]\}~,
     \ee
     where $E^{\pm}=E(\frac{1\mp\gamma_5}{2})~,~~E=\frac{1}{2}g\rho
     K$,  and the operator $K$ is defined by Eq.(1.6). In order to
     get the effective  chiral Lagrangian one has to integrate (5.3)
     over all meson fields exept for the Goldstone octet $\pi_b$.
     Using the steepest descent  in the vicinity of the vacuum
     sadde--point [1,4] we get
     $$
     S_{eff}(\pi_b)=-\ln \int D\xi D\mu D\varepsilon \tilde{Z}_{\psi}=
     $$
     \be
     =tr\ln(i\hat{D}+iEU_5)~,
     \ee
     where
     \be
     U_5=U(\frac{1-\gamma_5}{2})
     +U^+(\frac{1+\gamma_5}{2})=\exp\{\frac{-i1}{2}\pi_a\lambda^a\gamma_5\}~.
     \ee
     The form of the effective action (5.4) coincides with that found
     in [1] for $N_f=2$. In absence of background gluonic field
     $B_{\mu}$ the form (5.4) reduces to the effective action found
     in [4] for the pure instanton case (see also discussion in
     [11]).

     The authors wish to express their gratitude to D.I.Dyakonov and
     V.A.Novikov for helpful discussion and to D.Kuzmenko for his
     contribution to the Appendix.

The final note is that the authors became aware of Ref. [5] after
 completing the present paper.
 Although our approach differs from that of [5],
 the results we got for
 $N_f=3$ fit into the sheme outlined in [5] for arbitrary $N_f$.

 \setcounter{equation}{0}
\renewcommand{\theequation}{A.\arabic{equation}}

\begin{center}
{\bf APPENDIX}
\end{center}

In the Appendix we show how the result (3.9) emerges. Consider the
integral (3.8):
\be
\tilde{\tilde{Z}}=\int D\sigma exp\{\int d^4u(2L\sigma
-\frac{\sigma^3}{\kappa^2})\}~.
\ee
We introduce a new variable $w$ and a new parameter $t$ according to
\be
w=\left(\frac{2\kappa^2L}{3}\right)^{-1/2}\sigma~,~~t=2
\left(\frac{2}{3}\right)^{1/2}\kappa
L^{3/2}~.
\ee
Then (A.1) takes the form (up to a Jacobian which is not essential
for our present purposes)
\be
\tilde{\tilde{Z}}=\int Dw exp\{\int
d^4ut(w-\frac{w^3}{3})\}~,
  \ee
  which is the standard representation of the Airy function [10]. The
  leading contribution (3.9) is then obtained by the steepest descent
  method. The integral has two  stationary points $w=\pm 1$, from
  which only $w=-1$ contributes. The integration contour can be
  choosen along the line $Rew=-1$.

     \end{document}